\documentclass[sigconf]{acmart}
\AtBeginDocument{%
  }

\setcopyright{none}
\acmDOI{}
\acmISBN{}
\acmConference[LIMITS '26]{12th Workshop on Computing within Limits}{June 23--25, 2026}{Online}

\begin{document}

\title{Modest, artistic, and radical solutions to the environmental impact of image-generating machine learning }


\author{Laura U. Marks}
\affiliation{%
  \institution{Grant Strate University}
  \city{British Columbia}
  \country{Canada}}
\email{lmarks@sfu.ca  }

\author{JESS MACCORMACK}
\affiliation{%
  \institution{Simon Fraser University}
  \city{British Columbia}
  \country{Canada}}
\email{jam47@sfu.ca}

\author{Kehui Li}
\affiliation{%
  \institution{Simon Fraser University}
  \city{British Columbia}
  \country{Canada}}
\email{kehui_li@sfu.ca}

\begin{abstract}
Machine learning is often touted to improve the efficiency of ICT, but that small gain is overwhelmed by the enormous carbon, water, and land footprints of data centers and ML-ready devices. We survey the electricity consumption of ML applications in training and inference, focusing on electricity-intensive image generation. Our team of a computer engineer, a media scholar, and an artist explore solutions including inexact computing; tiny language models; low- precision hardware architectures; hardware with limited capacity; and anticipating and mitigating energy demands at the design phase. We will sketch our work in progress of an ethical and aesthetically sophisticated tiny image generator using non-scraped data. Looking to the economic context, we will propose a true-cost accounting for the environmental impact of machine learning and suggest that the criterion of efficiency is driven by the shareholder-capitalist framing of ICT.

\end{abstract}



\maketitle

\section{Background}
Users are encouraged to consume image-generating AI as a fun pastime with no material impact. For a sense of the environmentally devastating consequences of this attitude, here’s a thought experiment. In the Mission: Impossible franchise, a mysterious Entity capable of generating universal and instantaneous deep-fakes threatens to take over the world. The Entity would need to use some combination of high-definition virtual reality and image-generating ML. Bracketing for a moment the environmental impact of training and the production of devices, let’s try to measure the energy consumption of the user’s device and estimate its carbon impact. At the user’s end, a single image generation generates 300 grams of CO\textsubscript{2}e \cite{luccioni2024power}. That suggests video generation, at a framerate of 15 fps (not 30 fps, accounting for the efficient use of keyframes), would generate 4500 grams of CO\textsubscript{2}e per second. If we say the Entity needs to fool one person per day with just 10 seconds of generated video, that’s 45 kilograms of CO\textsubscript{2}e. A larger-scale hallucination, say on the level of 100,000 people, would produce 4.5 megatons of CO\textsubscript{2}e. So it appears that this brief collective hallucination would generate in ten seconds as much pollution as 1.2 coal-fired power plants would in a year \cite{epa}. 

Basically, for the Entity to achieve its nefarious ends, half the planet would need to be paved with data centers, power stations, solar panels, and mineral extraction sites. Before it succeeds in its evil mission, we would all have choked to death. Yet such scenarios are blithely presented to publics as being on their way to actuality. 

ICT is demanding ever more and larger data centers to store enormous files and crank massive calculations. The networks that transmit those files and instructions, and the user-end devices that run them also use ever more electricity, in their use and in their production. ICT’s increasing electricity use is part of increased electricity demand over all, especially in wealthy countries. Some of these increases appear to result from decreased use of fossil fuels \cite{morley2018digitalisation}; but ironically, that electricity comes mostly from fossil fuels, currently the source of 80\% of global electricity. All energy production is expanding—renewables and fossil fuels. Energy efficiencies would most likely be outweighed by corporations’ determination to grow their markets \cite{pihkola2018evaluating}. In the rebound effect, increasing efficiency of ICT infrastructure and devices has made possible huge expansions of energy-intensive applications like AI, crypto, online gaming, and high-resolution streaming, which remains still one of the main drivers. 

AI data centers are drastically expanding ICT’s contribution to global warming. Projected to use five times the electricity of traditional data centers \cite{geiger2023datacenters}, data centers are expected to reach 35 gigawatts of power consumption annually by 2030, up from 17 gigawatts in 2022 \cite{mckinsey2023investing}. The electricity use of data centers in the United States more than doubled from 2018 to 2023, from 1.9\% of total US electricity to 4.4\% in 2023, and they project that it will consume 6.7\% to 12\% in 2028 \cite{shehabi2024us}. IT and specifically machine-learning companies are building power plants at an extraordinary rate to support the surge of energy needed to power data centers that run ML. As important as greenhouse gases is the water footprint, mostly for cooling data centers. Many AI data centers located in areas that already face water shortages. Google, Microsoft, and Meta’s data centers used 2.2 billion cubic meters of water in 2022—half the annual water usage of the UK—mostly for AI \cite{li2024making}. 

Corporate speculation drives this expansion of infrastructure. In ICT’s vicious cycle, infrastructure is expanded in anticipation of needs; “needs” are created, to justify the expansion of infrastructure; and infrastructure expands further. We’ve seen this pattern already with online video. In November 2025, the bank JP Morgan Chase predicted that by 2030, companies would invest 5 trillion USD worldwide in data center infrastructure and power supplies. The data centers would use 146.4 gigawatts of power \cite{moss2025jpmorgan, wigglesworth2025global}. 

It is certain that total energy supply will decrease drastically. Renewables are only meeting the increased demand for electricity, or at best slightly above it. \cite{sgouridisSowersWayQuantifying} finds that to make the transition to renewable energy, in order to meet the IPCC recommendation to keep planetary temperature rise below 1.5 degrees, would require acceleration of renewable energy installation at 30-50 times the 2016 rate. This is a gloomy prospect.

In reviewing this literature, we have noticed that AI designers, even those not committed to environmental computing, often exhibit good values. We note the pride that computer engineers and architects take in their work e.g. \cite{deng2021lowprecision}, as well as the value of open-source generosity that many exhibit. The demand to develop more electricity-intensive applications arises from the corporations that benefit from this work. 

We echo Knowles et al. [2022]’s call for an end to digital exceptionalism, whereby ICT gets a pass on environmental issues because its products are supposedly so valuable. A systems approach shows that when rebound effects are taken into account, the only sustainable path forward is degrowth \cite{hickel2020less, santarius2023digital, widdicks2023systems}. Sustainability researchers and makers call for the undesign of ICT to remove those aspects of digital media that consume untenable levels of resources and energy \cite{pierce2012undesigning}. This includes eliminating technologies like “non-fungible tokens, cryptocurrency, large language models, immersive virtual environments, on-demand streaming platforms, and others with higher resource and energy consumption” \cite[p.~20]{nardi2018computing}. If any version of LLM and generative imaging applications survive this reasonable constraint, they will have to be modest, streamlined, and useful for purposes other than profit. 

\section{Electricity consumption}
Market-predicting institutions like the International Energy Agency \cite{iea2025energy} assume that AI’s requirements for energy and infrastructure are a fait accompli and that energy providers can only react to predicted demand. Though estimates vary, most predict a dramatic jump in consumption. 

\textit{Circuits}. The electricity cost of machine learning lies mainly in circuits, training, and inference. First, the circuits that support AI and other energy-intensive applications are proliferating rapidly in data centers and user-end devices. AI can be processed on regular CPUs, but accelerators, mainly graphics processing units or GPUs, allow the server to quickly process high numbers of calculations in parallel. Accelerators are used for the “wild card applications” of training and inference of ML models, the Internet of Things, virtual reality, cloud gaming, and blockchain, so called because they can cause huge spikes in energy consumption \cite{cisco2020annual}. \cite{shehabi2024us} notes that growth of US data centers began in 2017 due to increased use of GPUs to accelerate servers for ML. Servers are shipped with 2, 4, or 8 GPUs. NVIDIA held the top share of the market in GPUs until recently, with its GH200 and earlier chips; others include AMD’s MI 350, Intel’s Gaudi 3, AWS’s Trainium3, Alphabet’s Trillium, Alibaba’s ACCEL, IBM’s Northpole, and Huawei’s Ascend910C. The latter is likely the chip used in DeepSeek both for training and for inference on user-end devices \cite{dilmegani2025top}. NVIDIA now markets a 72-GPU server.

Then there are edge-computing chips designed to carry out AI on users’ devices: their manufacturers include NVIDIA, Google, and Intel. Although most of the analysis of AI’s electricity consumption focuses on data centers, it’s important to note that user-end devices increasingly act as mini-servers, as the calculation work of ML is passed on to the user’s computer or phone. In edge-server computing, data are sent from the device to the edge server for computation. In fog computing, the processing occurs at the LAN (local-area network) level \cite{merenda2020edge}. Telecoms around the world are building new data centers to support fifth-generation or 5G computing, which relies on core and edge network servers, moving computational tasks closer to the user end. 

\textit{Training ML}, especially large language models, requires repetitions of calculations using billions of parameters. Training a single AI model can emit as much carbon as five cars in their lifetimes \cite{strubell2019energy}. Accuracy is generally prioritized over efficiency in LLM training. But increased accuracy has diminishing returns when the cost of training is included. Deep learning progress depends on increasing computing power. As parameters increase, flexibility and accuracy increase, and the ability to improve performance also increases \cite{thompson2021computational}. As Thompson and colleagues point out, when the setting is overparameterized, computational requirements grow at least as the square of the number of data points. As they estimate, the computation required to train an overparameterized model should grow at least as a fourth-order polynomial with respect to performance. They estimate that training an ImageNet model to an accuracy of 5\% would produce almost as much carbon as a medium-sized city in a month, while an accuracy of 3\% would require as much carbon as New York City emits in a month \cite{thompson2021computational}. In another example, the BLEU (bilingual evaluation understudy) score rates accuracy of an ML translation model. A neural architecture search achieves a new state-of-the-art BLEU score of 29.7 for English to German machine translation, an increase of just 0.1 BLEU at the cost of at least \$150k in on-demand compute time and non-trivial carbon emissions” \cite{so2019evolved} \cite[p.~4]{strubell2019energy}.  

Given the high ratio of increased power to increased accuracy, it is highly preferable to aim for models that are just sufficiently accurate. 

The gain in energy consumption would be driven mostly by AI inference—the power used when interacting with a model—rather than AI training. Increasingly people use ML apps for tasks that a calculator or search engine (or one’s own brain) could do, at much higher electricity cost. Individual inferences use much less computation, less electricity, but popular models run millions or billions of inferences per day, e.g. Google Translate, ChatGPT, or Google’s switch to AI searches \cite{luccioni2024power, smith2025hidden}. According to \cite{smith2025hidden}, based on OpenAI reports generation-AI ChatGPT has 700 million weekly users and serves more than 2.5 billion queries per day. In a blog post OpenAI CEO Sam Altman’s estimate of 0.34 watt-hours per query. If an average query uses 0.34 Wh, that’s 850 megawatt-hours per week; enough to charge thousands of electric vehicles every day.

\textit{Inference}. The gain in energy consumption would be driven mostly by AI inference—the power used when interacting with a model—rather than AI training. Increasingly people use ML apps for tasks that a calculator or search engine (or one’s own brain) could do, at much higher electricity cost. Individual inferences use much less computation, less electricity, but popular models run millions or billions of inferences per day, e.g. Google Translate, ChatGPT, or Google’s switch to AI searches \cite{luccioni2024power, smith2025hidden}. According to \cite{smith2025hidden}, based on OpenAI reports generation-AI ChatGPT has 700 million weekly users and serves more than 2.5 billion queries per day. In a blog post OpenAI CEO Sam Altman’s estimate of 0.34 watt-hours per query. If an average query uses 0.34 Wh, that’s 850 megawatt-hours per week; enough to charge thousands of electric vehicles every day.

Luccioni and colleagues measured electricity consumption of inference and CO\textsubscript{2} emissions of different types of machine learn- ing/AI per inference \cite{luccioni2024power}. They calculate that text classification generates 0.4g CO\textsubscript{2}  per action. Text generation, like ChatGPT generates 9g CO\textsubscript{2}  or about 20x text classification. And image generation produces 300g CO\textsubscript{2} , or 750 times that of text classification. So an artist, designer, or malicious deepfaker can be working quietly in their studio while simultaneously generating a massive carbon footprint!

\section{Toward tiny machine-learning models}
ML designers and researchers prioritize computationally efficient hardware and models. However, the goal is mostly to increase speed and accuracy, not to decrease electricity use. For example, parallelism, dividing tasks among computing devices, is efficient, but only saves time, not energy. The criterion of efficiency has dubious value. Newer models are not necessarily more efficient; they’re more complicated. Training on new hardware can be faster, but faster processors use more power. Exemplifying the rebound effect, demand for ever more sophisticated ML models leads to greater efficiency in order to do even more. What is needed is self-sufficiency \cite{hilty2015computing} or digital sufficiency \cite{santarius2023digital}. As we note later, self-sufficiency is barely credible in the current market, but it might be different if the cost of climate damage were included in the cost of ML model training and use in a true-cost accounting. All but the most truly efficient models that demonstrably serve important social goods would dwindle, perhaps to be used only on an inexorably priced subscription basis. 

\subsection{Current solutions for low-energy image-generation models}
\subsubsection{Lower precision} 
\label{sec:subsubsection}
Approximate or inexact computing, or computing that solves to fewer decimal points \cite{barua2019approximate}, offers a host of methods to reduce the energy and computational time required for tasks that not need accurate but “good enough” results. low- precision hardware architectures store parameters in low-precision formats. Quantization decreases size but increases error. Deng and colleagues at Facebook \cite{deng2021lowprecision} work on 8-bit and 4-bit quantization, which incurs problems when working with proprietary software. The authors share their tools open source. Given Facebook ’s over- whelming need for speed, very small (or low-bit) LLMs do not meet their needs—but they may for other tasks. Ironically, some of the changes they hope for would require more memory, either larger on-chip memory or more bandwidth, meaning that to make the ML faster would require more electricity and result in higher emissions. In contrast, \cite{xu2024onebit} show that 1-bit parameter representation of LLMs is pretty good for common-sense reasoning and world knowledge.

As usual with industry-driven developments, the goal of efficiency is not to use less energy (more with less) but to do more with the same amount of energy, and ultimately, given the slippery slope of the rebound effect, to do more with more energy. Xu and colleagues explain that the goal of their research, since high-bit models not accessible on smaller devices, is to make LLM available on devices with 1 GPU. This would lead to more use of LLMs, more GPU-equipped devices, and more electricity use, in another case of the rebound effect. 
Avoiding unnecessarily large tools Comparing neural to non- neural models of text classification, \cite{cunha2021cost} found that non-neural approaches worked best when data was limited, but for large amounts of data neural approaches performed better. Shifting to non-neural approaches produced a more than 23× speedup with a 5\% drop in performance.

\subsubsection{Small language models} 

(SLMs) are ML models that would use a very few parameters for a specific task. Small language models, machine-learning applications like TinyGPT-V, use “few-shot” algorithms and parameter counts that are degrees of magnitude smaller than those of LLMs \cite{schick2021small}, SLMs can run on a single GPU, using relatively little electricity in training and use \cite{thomas2023small, wang2024small}. General intelligence is unsustainable by any measure; the goal of these projects is specific intelligence. 

\subsubsection{Small hardware} 
Some ML is designed small so that it can be deployed on hardware with limited capacity, such as Raspberry Pi, Arduino, SparkFun Edge, and AdaFruit EdgeBadge \cite{merenda2020edge}. The initial purpose of this is to allow millions of edge devices to run ML for IoT applications, and thus it is another example of rebound. However, useful workarounds developed for edge devices include smaller datasets; pruning, i.e. using fewer weights \cite{merenda2020edge}; and sparse tree algorithms like Bonsai \cite{gope2019ternary}.

\subsubsection{Anticipate and mitigate the energy demands in the design phase} 

Some researchers \cite{faiz2024llmcarbon} recommend that LLM designers accounting for parameter count, hardware configurations, and energy efficiency of the data center hosting the model, so that designers can “intelligently explore the trade-off between test loss and carbon footprint when designing new LLMs.” Keep hardware longer—but not too long. Faiz and colleagues’ \cite{faiz2024llmcarbon} equation for calculating the embodied carbon footprint of LLM, the operational energy (energy-oper) associated with LLM processing, divides the sum by lifetime of the hardware. The longer you use it, the lower its environmental impact will be. More efficient hardware, i.e. servers, will come along, but it’s a big question whether the gain in efficiency would offset the embodied carbon of the server. However, \cite{andrae2023large} argues that since the large proportion of electricity consumption is at the user’s end, and newer devices are more electricity-efficient, it is crucial not to keep ICT network infrastructure for more than 6 years. 

\subsubsection{Remove corporate-driven criteria for speed and accuracy} 

\cite{deng2021lowprecision} insist that low-precision models must be no more than 5\% less accurate than the full-precision model. P99 latency statistical measure of response time. So 100-ms p99 latency means 99\% of responses occur within 100 microsecond, 1/100,000 of a second. This is a nutty criterion, driven by corporate need for eyeballs not to stray from Facebook. It’s an addiction-driven model. If we lower expectations to a reasonable 1/10 of a second from user’s query to response, the model will shrink to a relatively tiny size. 

\section{Rebound effect redux}
ICT engineers are working so hard to make ML models more efficient. But again, efficiency is destined to generate yet more use of ML. \cite{xu2024onebit} note that high-bit models are not accessible on smaller devices, and that their goal is to make LLM available on devices with 1 GPU. That is, phones, as well as computers, equipped with GPUs will be able to run LLMs. This would lead more consumers to buy new phones and gaming devices equipped with GPUs (such as the Apple A16 Bionic and A17 Pro, the Snapdragon 8 Gen 2 used in Samsung Galaxy S25 Ultra and other phones, Huawei’s Kirin 9000S, PlayStation 3 and higher, and Xbox: devices that arrive from the factory with a whacking embodied carbon footprint. Consumers would use their GPU-equipped devices mostly unwittingly, in edge computing and to decompress high-resolution video on smart TVs, as well as intentionally to run AI models. Collectively their devices would consume megawatts of electricity. The phones will likely tire after a couple of years, and even if they do not, companies will devise compelling reasons (perhaps such as fitting two GPUs on a device) for consumers to ditch this phone for a new one. Unless the cost of electricity is revised to reflect its carbon footprint, which right now seems as likely as me growing a third eye, more efficient ML models will be burning more coal in a matter of months. 

Lange and colleagues compare four effects of digitalization on energy consumption: the direct effects of ICT production, use, and disposal; ICT-led energy efficiency increases; economic growth; and tertiary effects. The authors use an extensive literature survey and analytical modeling. They conclude that overall, digitalization increases energy consumption. Lange et al. find that the majority of researchers conclude that rebound effects are great enough to prevent a sufficient absolute reduction in energy demand” \cite[p.~5]{lange2020digitalization}. 

In this and other studies, the problem with estimating rebound effect is that it is difficult to model user behavior and the resulting rebound effects. In the case of ML, this would require studies of how often consumers in different regions use various kinds of LLMs, intentionally or not, keeping in mind the variation in electricity demand of different models (as we’ll see). 

\section{Design of a tiny image generator}
We considered all the possible methodologies for building an ethical, tiny image generating model as a conceptual artwork that would contrast with the accelerationist goals of LLM to absorb human knowledge and art and then privatize the models for profit. For us, this would begin by training a model on ethically sourced images. Rather than working with a scraped database, we would invite participants to submit their own images as initial artworks. We considered what would compel them to be part of this project and decided that the model would reflect the community that uses it and recognizes their contributions. Thus, participants’ metadata must be treated as part of the artwork and published. 

We are inspired by the Slow AI project of AIxDESIGN, a year-long collaborative project by critical ML researchers, designers, technologists, and artists. Questioning the demand that AI be scalable and general and that scraped data is acceptable, they work with small-scale datasets, data sovereignty, process-oriented approaches, and localized functions \cite{flynn2025aixdesign}. 
 
Together we tried to imagine a possible output that would prompt the users to consider their role in the language model as well as possible socio-political repercussions of AI use. What kind of imagery could be created from a very small dataset? We considered the ways LLMs, because of their reliance on statistical rendering, tend to generate “mean images,” in Steyerl’s term \cite{steyerl2023mean}. Mean images, rather than abstract from an existing phenomenon, generate the phenomenon (like Googling “Iran and terrorism”). Largely because of the scraped data on which they rely and the heavily clich\'ed patterns they reinforce, mean images “pick up on latent social patterns that encode conflicting significations as vector coordinates” \cite[p.~18]{steyerl2023mean}.

Instead, we wanted iterations to be novel, randomized and based on elements of surprise. In activation, the model would learn complex real-world patterns and information. Activation function enables the model to learn and represent complex data patterns. The activation function would determine what kind of information or pattern we want the model to learn and represent \cite{pavlov2025controlling}. 
Different activations produce different styles, and the parameters control specific style aspects. As Pavlov states, “Slight changes to the parameters of these functions will lead to slight changes in the resulting image, while larger changes lead to more significant and unpredictable alterations to both the structure and style of the image” \cite[p.~4]{pavlov2025controlling}. 
(We continue in the subjunctive, as we may not be able to carry out the project. Computer scientist Stephen Makonin helped us a great deal at the outset but was unable to continue.) 

\subsection{Approaches for model size reduction}
\subsubsection{Hardware} We would train the model on a Raspberry Pi computer. 
\subsubsection{Few-, One-, or Zero-shot training algorithms} These are ma- chine learning techniques in which a model learns to recognize new objects or patterns from a single example or a very limited number of examples \cite{schick2021small}. They do not directly make the model smaller, but they make training more data-efficient. One-shot training algorithms are usually used when there are not many examples in the training set, allowing the model to generalize from a smaller dataset. Still, we would need a lot of images to train a model, as well as time and resources to add metadata.

\subsubsection{Quantization} This is a technique that reduces the precision of numerical values in a model. Simply, it uses fewer bits to represent the model, which makes the model size smaller. For example, mixed- precision quantization uses multiple precision levels to represent different parts of the model. Important layers use higher precision to preserve accuracy, while less critical parts use lower precision. This reduces model size while maintaining overall accuracy. of weight matrices \cite{deng2021lowprecision}.

\subsubsection{Noise} The model would noise to input images rather than working with averages. To add noise in an interesting way, we would assign different weights to the datasets. We would use the StableDiffusion v2-1-unclip (small) Model Card from HuggingFace \cite{rombach2022highresolution}. StableDiffusion uses LAION-5B, a relatively ethical, publicly available dataset derived from Common Crawl data scraped from the web. While Stable Diffusion models are considered relatively small compared to industry giants like FLUX (which has up to 12B parameters), the SD 2.1 unclip-small variant is specifically the “small” member within its unCLIP architecture family. At approximately 1.2 billion parameters, it is not particularly small when compared to truly lightweight models like DreamLite (0.39B) or Nitro-E (0.3B). Furthermore, as a fine-tuned version of its parent model (SD 2.1), the emissions from training the parent must be accounted for. This results in a total carbon footprint of roughly 15,000 kg CO\textsubscript{2} equivalents. 

We would adapt StableDiffusion to add noise and surprise to image-plus-word prompts, heavily weighted with our own by-permission image database. In this model the amount of noise added to the image embedding can be specified, from 0 for no noise to 1000 for full noise. We would experiment (but not too much) with higher levels of noise.

\subsection{Artistic approaches to tiny image-generation models}
    
\subsubsection{Novelty} The goal of this tiny image generator is novelty, not averages, and to magnify artifacts. Users would input an image and receive a surprising result. As an art project, the goal is not to generate images with verisimilitude but to create interesting variations on the input images that tell us something about the model. By responding to previous input images, the model will start to reflect the community that uses it. With participants’ permission, we would to publish the metadata the project generates as a way of acknowledging their input. Ideally we would make the model available as an art kiosk.
	
\subsubsection{Dataset choice} To augment the LAION-5B dataset on which StableDiffusion is trained, we contacted the Victoria \& Albert Museum about donating a dataset of textiles and Islamic calligraphy from their extensive archives. Heavily weighted, these images would add interesting texture and content. The model would prioritize learning the dataset’s visual features—textures, geometric patterns, arabesque, and calligraphic forms—over other styles or the pre-trained model’s existing biases. We asked, What other ethically sourced databases could we create or buy to add to the model? What is the smallest dataset we could work with successfully?

\subsubsection{Dataset weighting} This is a technique where different data sources are assigned different levels of importance during model training or fine-tuning. Instead of treating every training example equally, we would explicitly tell the model which examples matter more.

There are three key problems that dataset weighting solves for our project. First, without weighting, a small dataset like ours would have minimal influence compared to larger datasets or the model’s pre-trained knowledge; weighting amplifies its influence, ensuring its features are learned thoroughly. Second, pre-trained models already contain strong visual biases toward photorealism, Western art, and other common styles; without weighting, these existing styles would dominate and dilute the Islamic art features we want to emphasize. Third, when multiple styles are present in training data, unweighted training causes all styles to blend equally, producing unpredictable and generic results. Weighting gives us explicit control over the blend ratio, allowing us to create intentional, controlled hybrids where Islamic art dominates while other styles provide subtle contrast.

\subsubsection{A political element} Most AI has been developed in the global North and therefore reflects hegemonic interests. This directly impacts Palestinians as the dehumanizing and Islamophobic portrayal in media and by AI allows for the continuation of the genocide. Israel makes use of and is invested in AI-based facial and emotion recognition technologies, automated weapons, social media monitoring, and military targeting, and it has used machine learning to generate targets for bombing in Gaza \cite{kawashImpactsAITechnologies, kwetHowUSBig}. In this light we decided to add one category that is always required to be used in the final output: a reference to Palestine, such as the name or an image of a watermelon. Participants’ input would constitute another small dataset. 

The resulting images would be limited in scope, elegant, funny, and decolonial. We would learn to appreciate austerity as an aesthetic quality of creativity within limitations.

\subsection{Reflection}

Shareholder capitalism drives the astonishingly fast rise of ML. In late 2025 the value of the world’s top ten companies was \$25.6 trillion. \$15.1 trillion of that amount had accumulated since November 2022, directly related to machine learning/AI \cite{lanchester2025king}. An effective, if unpopular, solution to the environmental impact of machine learning would be if governments price electricity for tech companies according to its environmental cost. Then tech companies would really push to be more efficient. They would, of course, pass costs on to consumers, so we’ll use those platforms more mindfully. 

In the shareholder-capitalist economy, efficiency leads to doing more with more. Efficiency in itself is thus not a conservationist force but a driver of electricity consumption. As Hilty points out, competition among software providers is one of the drivers of obsolescence \cite{lambert2015postpeak}. Similarly, Lange and colleagues argue that “Drastic absolute decoupling [of economic growth from energy consumption] is necessary to remain within planetary boundaries” \cite{lange2020digitalization}. But such decoupling is not feasible in an economy that defines planetary heating as an externality. 

In contrast to the market-driven, planet-burning boosterism of ML-led corporations, we advocate the models of digital sufficiency \cite{hilty2015ict, santarius2023digital}, collapse informatics \cite{tomlinson2013collapse}, and graceful degradation \cite{lambert2015postpeak} as the only ways ICT can contribute to environmental healing rather than increase harm. 

We hope media makers and users will simply eschew generative imaging, as it typifies digital excess \cite{olssonUserCenteredLensDigital} and the cornucopian paradigm \cite{preistUnderstandingMitigatingEffects, widdicksBreakingCornucopianParadigm}. But for people who are required to work with generative imaging, for example, in their jobs, this model shows that it is possible to create that uses much less energy in both training and inference. The approach we propose can also inform policy.

Enticements to opt in to lower quality include the awareness of AI’s unsustainability; voluntary simplicity \cite{etzioniVoluntarySimplicityCharacterization}; the desire to make convivial tools \cite{vetterMatrixConvivialTechnology}; and an interest in creative experimentation. Our approach is inspired by the Small File Media Festival, which invites makers to create videos that stream at no more than 1.44 MB/minute \cite{smallfileSmallFileMedia}. We will promote best practices for tiny generative imaging of the widely consulted SFMF website.

\bibliographystyle{ACM-Reference-Format}
\bibliography{paper-base}

\end{document}